\documentclass[sigconf, authorversion]{acmart}
\usepackage{mathtools} % This package enhances math typesetting  
\usepackage{algorithm}
\usepackage{algpseudocode}
\usepackage{balance}

\AtBeginDocument{%
  }

\copyrightyear{2026}
\acmYear{2026}
\setcopyright{cc}
\setcctype{by}
\acmConference[KDD '26]{Proceedings of the 32nd ACM SIGKDD Conference on Knowledge Discovery and Data Mining V.2}{August 09--13, 2026}{Jeju Island, Republic of Korea}
\acmBooktitle{Proceedings of the 32nd ACM SIGKDD Conference on Knowledge Discovery and Data Mining V.2 (KDD '26), August 09--13, 2026, Jeju Island, Republic of Korea}
\acmDOI{10.1145/3770855.3818355}
\acmISBN{979-8-4007-2259-2/2026/08}

\begin{document}

\title{Deep-learning Causal Retrieval Optimization for Efficient e-commerce Distribution in Pinterest}

\author{Junpeng Hou}
\email{jhou@pinterest.com}
\affiliation{%
  \institution{Pinterest, Inc.}
  \city{San Francisco}
  \state{CA}
  \country{USA}
}

\author{XianXing Zhang}
\email{xianxingzhang@pinterest.com}
\affiliation{%
  \institution{Pinterest, Inc.}
  \city{San Francisco}
  \state{CA}
  \country{USA}
}
\authornote{Work performed at Pinterest, Inc.}

\author{Sai Xiao}
\email{sxiao@pinterest.com}
\affiliation{%
  \institution{Pinterest, Inc.}
  \city{San Francisco}
  \state{CA}
  \country{USA}
}

\author{Derek Cheng}
\email{naichiacheng@pinterest.com}
\affiliation{%
  \institution{Pinterest, Inc.}
  \city{San Francisco}
  \state{CA}
  \country{USA}
}

\author{Darren Reger}
\email{dreger@pinterest.com}
\affiliation{%
  \institution{Pinterest, Inc.}
  \city{San Francisco}
  \state{CA}
  \country{USA}
}

\author{Olafur Gudmundsson}
\email{ogudmundsson@pinterest.com}
\affiliation{%
  \institution{Pinterest, Inc.}
  \city{San Francisco}
  \state{CA}
  \country{USA}
}

\author{Mehdi Ben Ayed}
\email{mbenayed@pinterest.com}
\affiliation{%
  \institution{Pinterest, Inc.}
  \city{San Francisco}
  \state{CA}
  \country{USA}
}

\author{Zhiqing Rao}
\email{zrao@pinterest.com}
\affiliation{%
  \institution{Pinterest, Inc.}
  \city{San Francisco}
  \state{CA}
  \country{USA}
}

\author{Huizhong Duan}
\email{hduan@pinterest.com}
\affiliation{%
  \institution{Pinterest, Inc.}
  \city{San Francisco}
  \state{CA}
  \country{USA}
}

\renewcommand{\shortauthors}{Junpeng Hou et al.}

\begin{abstract}
Pinterest is where people turn inspiration into action as users browse ideas, then take steps toward realization, often by discovering shoppable content. To support this journey, we must distribute commerce content when it helps, not when it distracts. We frame this as a causal decision of triggering shopping candidate generators in early retrieval and deploy a production system at Pinterest that learns personalized and contextualized triggering policies. A deep multi-task model jointly predicts outcomes and uplift of multiple events, trained with a doubly-robust pseudo-outcome alongside calibrated outcome losses for stable, single-robust uplift learning. A randomized data logging supplies counterfactual coverage, and the model is evaluated by both regular and reverse metrics for full assessment. A linear-time offline replay is designed to select thresholds and forecast policy impact with extremely high consistency with online results.
For productionization, the model runs in parallel with remote retrieval calls without end-to-end latency regression. At web scale, we cut shopping triggers by up to 85\% while holding key shopping sessions neutral, improving important total sessions (+0.26\%) and Pin saves (+1.10\%), with significant infrastructure savings.
By unifying deep causal learning with reliable offline replay and demonstrating production-grade deployment, this work provides a generally practical recipe for early-retrieval optimizations in modern cascading recommenders beyond shopping, aligning exploration and cost with user intent at scale.
\end{abstract}




%   \includegraphics[width=\textwidth]{closeup_exp.png}
%   \caption{Recommendation funnel and visual shopping journey in Closeup.}


\maketitle

\section{Introduction}
Pinterest is the visual discovery platform where people come for inspiration and then take steps toward realization, often by discovering and acting on products that match. As a result, shopping distribution plays a pivotal role in translating ideas into action and commerce content should appear when it helps the journey, not when it interrupts exploration. Our recent efforts on embedding-based retrieval in Closeup substantially improved the quality and breadth of shoppable content \cite{Hou2025Optimize}, but they also sharpened a core product question: for any given request, when should we distribute shoppable content to users?

Answering ``when to trigger'' is consequential as indiscriminate exposure to shopping content can backfire. In our logs, uniform shopping triggering lifts shopping actions like visiting merchant webpage but simultaneously depresses non-shopping engagement like saving Pins and social events, a direct evidence of a delicate trade-off between commerce and inspiration that cannot be resolved by static rules or downstream fixes alone. Moreover, at Pinterest scale, this decision is made billions of times every day, under tight latency budgets and in the presence of competing objectives across sessions and surfaces. This motivates an intent-aware approach that reasons about incremental impact, not just correlation, and that can personalize the triggering to different users and contexts. To address this, we need to answer the following two critical questions:

i) How to frame the triggering decision as a machine learning problem and how to approach it on a large scale, like Pinterest?

ii) How to reliably assess triggering policies offline for fast iterations without affecting user experiences in production?

Led by the above arguments and the efforts in improving notification in social media \cite{InsNotif, PinsNotif}, we cast the triggering decision as a causal policy problem embedded directly in early retrieval. Concretely, we learn personalized policies that decide whether to fire shopping candidate generators (CGs) by estimating the incremental effect of triggering on business metrics, using a deep multi-task model that jointly predicts potential outcomes and uplift on multiple objectives. To support identification and robust iteration, we log randomized Shopping Holdout traffic and develop a linear-time offline replay procedure that forecasts policy effects and selects serving thresholds before launch. Together, the framework gives us a practical, trustworthy offline loop for a decision that historically required costly online experimentation. Our key contributions are threefold:

i) Formulation and productionization of causal trigger learning for early retrieval (see Fig.~\ref{fig:sys_overview} for a system overview), with a deep multi-task architecture and doubly-robust (DR) uplift training.

ii) A practical offline-to-online workflow with randomized Shopping Holdout plus linear-time replay that reliably selects thresholds and forecasts impact before launch.

iii) A web-scale deployment that delivers measurable engagement gains and cost reductions without latency regressions, along with stability and interpretability analyses that clarify how the policy generalizes.

By moving the ``when to trigger'' decision into a causal, intent-aware early-retrieval policy optimization and by making it learnable, auditable, and deployable, we provide a general recipe for modern cascading recommender systems to balance competing goals at scale with positive engagements and huge infrastructure savings, and it can extended beyond shopping to address more general questions including ``how to trigger''. 

\section{Related work}
\subsection{Causal inference}
Recent years have witnessed a steady proliferation of causal inference research and industrial deployments across large-scale recommender systems and digital platforms. Early industrial milestones set the foundation for counterfactual evaluation and learning in web environments, exemplified by the introduction and refinement of inverse propensity scoring and doubly robust estimators for unbiased offline evaluation at Microsoft Bing~\cite{graepel2010counterfactual, bottou2013counterfactual, li2015doubly}. Uber and Snapchat both adapted high-confidence off-policy techniques to the actual marketplace and feed optimization~\cite{grbovic2019high, kim2021improving}. Meta (Facebook) and Alibaba, meanwhile, developed production-ready experiment and recommendation frameworks to mitigate exposure and popularity bias through cluster-randomized tests and direct structural causal modeling~\cite{somani2020ab, liu2021eliminating}. Netflix and Amazon demonstrated the importance of causal embeddings and uplift modeling for improving candidate selection and robustness in recommendations~\cite{zhou2021causal, amazon2021causal}. More recently, a new generation of causal retrieval and triggering approaches has emerged: for example, YouTube’s Online Matching system uses contextual causal bandits for improved live exploration and recall~\cite{yi2023online}, and TikTok’s Douyin has implemented interference-aware causal estimators at production scale~\cite{farias2023correcting}. Modern academic contributions have further clarified that standard bias correction is insufficient, advocating for structured causal frameworks in RS~\cite{cavenaghi2023towards} and proposing model-free causal feature selection to directly enhance uplift in contextual bandits~\cite{zhao2024causal}. Together, these advances represent a convergence of theory and practice, embedding causal reasoning into the technological fabric of global recommendation and search systems.

\subsection{Causal policies for triggering decision}
Recent work explicitly treats the triggering decision—whether or when to fire a recall or recommendation—as a causal or bandit problem rather than a pure ranking task. Meta’s engineering team framed Instagram’s push notifications as a budgeted uplift problem, randomizing sends vs. drops of daily digests and training a neural uplift model to estimate each notification’s incremental effect on user activeness, then ranking by predicted causal lift and thresholding to meet a fixed send-rate, yielding large volume reductions with stable engagement \cite{Park2022Instagram}. Beyond notifications, two-stage recommendation studies show that trigger policies should be learned counterfactually with awareness of downstream ranking: Ma et al. propose an off-policy policy-gradient method that jointly optimizes the first-stage generator under logged data, demonstrating that ignoring generator–ranker interaction leads to suboptimal recall \cite{Ma2020OffPolicy}; at YouTube, Chen et al. deploy a large-scale REINFORCE recommender with off-policy and Top-K corrections to debias multi-item slate learning from logs, illustrating how counterfactual techniques enable practical trigger policies in web-scale settings \cite{Chen2019TopK}. In federated or multi-vertical settings, where dispatch chooses among generators and partial feedback plus cross-source competition complicate estimation, unbiased counterfactual evaluation of vertical-blending policies using logged propensities has been demonstrated on real multi-vertical search data \cite{Prochazka2019Vertical}, and production case studies show that interference at trigger time can severely bias naïve A/B estimates; interference-aware estimators such as Differences-in-Qs can recover nearly unbiased treatment effects in large marketplaces \cite{Farias2023Douyin}. Collectively, these advances replace rule-based or purely correlational intent gates with causal uplift models and contextual bandits for trigger/gating and multi-generator dispatch, increasingly validated through off-policy evaluation and interference-aware experimentation \cite{Park2022Instagram,Ma2020OffPolicy,Chen2019TopK,Prochazka2019Vertical,Farias2023Douyin,Ding2023FedCIRec}.

\section{Methodology} 
\subsection{Problem definition} \label{sec:problem_def}
To start with, we frame the problem as a binary decision to trigger the shopping CGs or not, given a user request $r$, which can be controlled by an incremental value or uplift
\begin{equation}
    \Delta_m(r) = p(m~|~y=1,r) - p(m~|~y=0,r),
\end{equation}
where $p$ represents the probability of hitting a successful metric $m$ given a triggering label $y$ and request $r$. Generally, the request $r = (r_u, r_c, r_q)$ contains all relevant information from user $r_u$, context $r_c$ and query $r_q$ if available. In this work, we assume that $y=1$ means that given CGs are triggered. In terms of choosing the successful metric, it can be any business-related reward ranging from simple counts of engagements to more complicated session metrics and retention indicators, like whether a user will be back. Here, we are mostly concerned with binary $m$ while it's easy to generalize to continuous rewards via expectation of returns $\Delta_m(r) = E(m~|~y=1,r) - E(m~|~y=0,r)$.

We can set the triggering decision by testing $\Delta>0$, which is referred to as the delta policy (DP) later. This policy simply states that the shopping CGs should be triggered when they could drive metric $m$ uniquely. More generally, we can define a personalized and/or contextualized threshold for triggering $\Delta>\delta_c(r)$ to allow more flexible controls over business objectives. One example is that we might want to onboard shopping content to new users more cautiously than heavy users. Throughout this work, we treat $\delta_c(r)$ as a constant for simplicity but also a tunable hyperparameter for fitting business goals.

However, the above uplift-based DP model can suffer from high variance in practice. The uplift score is the difference between two potential outcome estimates, so errors from both treatment arms can be amplified. In sparse-coverage regions, even the sign of $\Delta_m(r)$ can become unreliable, which makes direct deployment of DP risky despite its clean causal interpretation. Moreover, modern deep models require sufficient counterfactual coverage for all treatment arms, which could be expensive to build and maintain at scale.

Alternatively, we can move to the lower-variance end of the classical bias-variance tradeoff and rely only on $p(m\mid y=1,r)$. This is referred to as the single-value (SV) method in the following. SV does not directly estimate incremental value, but it is often more stable for production thresholding because it avoids subtracting two noisy outcome estimates. In this work, we study both methods: SV is used as the main production policy when offline-online consistency is critical, while DP serves as an important causal diagnostic for whether the uplift head has learned meaningful incremental signals.

\begin{figure}[h]
  \centering
  \includegraphics[width=\linewidth]{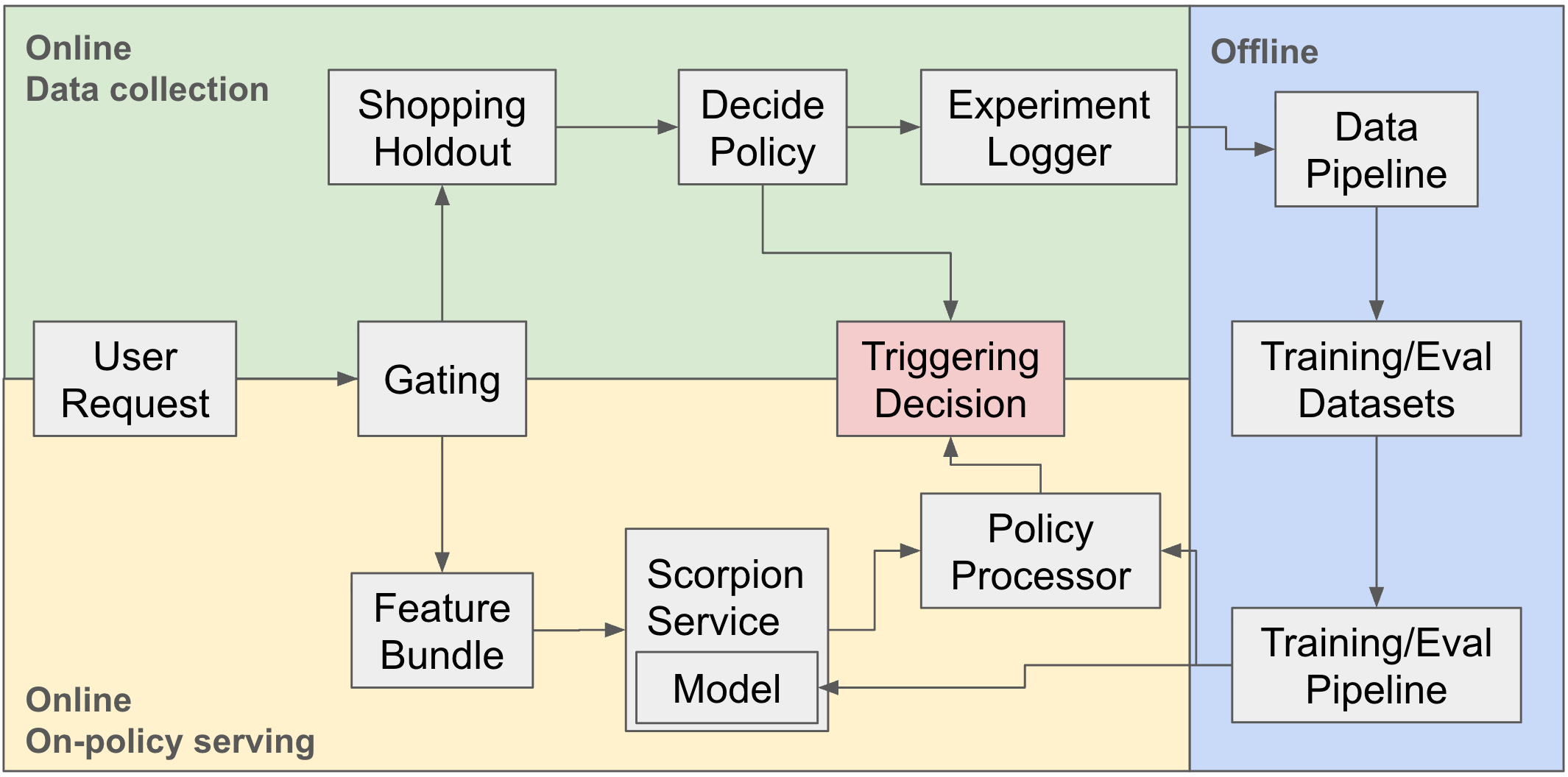}
  \caption{Overview of the whole system, including serving, logging and model training/eval.}
  \label{fig:sys_overview}
\end{figure}

\subsection{Data logging} \label{sec:data_logging}
To collect counterfactual data, a small fixed percentage of traffic is gated to a Shopping Holdout. Within this holdout, requests are randomized with equal probability between triggering and non-triggering, i.e., $p(y=1\mid r)=p(y=0\mid r)=0.5$ by design (see Online - Data collection part in Fig.~\ref{fig:sys_overview}). The assigned decisions are logged together with request features, and then passed into offline data pipelines that featurize the dataset and construct labels for training and evaluation (see Offline part in Fig.~\ref{fig:sys_overview}).

Although Pinterest operates at web scale, the effective counterfactual dataset for this problem is still highly sparse. Shopping content represents a small fraction of the overall ecosystem, and the randomized holdout is applied only on top of the shopping-eligible traffic (e.g., certain country filters apply). This makes the logged counterfactual table orders of magnitude smaller than standard ranking or retrieval logs. The controlled 50/50 design is therefore important because it guarantees overlap in the treatment assignment and removes the need for estimating sensitive propensities in the main experiments presented in the following sections.

While it's impossible to build exact counterfactual data for the same session, we could explore the full training data to find pairs of similar sessions with distinct triggering decisions. To define similar sessions, we can compare session-level features (e.g., nearest-neighbour) and compute the incremental value as the difference of some reward functions. Taking a step further, this can be viewed as, and generalized to, a value model at the retrieval stage. In this work, we find that it's not necessary to pursue perfectly counterfactual data since the logged data should already provide a good counterfactual coverage in users' feature space.

\subsection{Multi-task causal learning} \label{sec:loss_design}
We intend to build a deep-learning model to learn from the logged data, and we start with the learning objective.
Considering a mini-batch $B=\{r_i,y_i,m_i\}_{i=1}^N$, a straightforward way is to predict $\mu_i = p(m~|~y_i,r_i)$ and use a BCE loss to learn the score distribution directly
\begin{equation}
    \mathcal{L}_{score} = -\frac{1}{N}\sum_{i=1}^N \omega_i\left(m_i\log(\mu_i) + (1-m_i)\log(1-\mu_i)\right),
\end{equation}
where $\omega_i$ represents the weights for each record in the batch. While this model could achieve good prediction results in per-head evaluation, it usually shows worse performance on DP since it fails to learn the counterfactual aspects in the training data. 
% So in practice, we have to use SV with such a model and it basically simplifies the triggering task to identification of users who most unlikely to contribute to $m$ with $y=1$.

To better model the counterfactual data, we refine the learning objective by combining statistical methods from causal inference and deep learning modeling, like multi-task learning. For a given $m$, the model makes several predictions including both outcomes $\mu_i^{(y)} = p(m\mid y, r_i), y=\{0,1\}$, a propensity head $e_i =p(y=1\mid r_i)$ and an uplift head $\tau_i=\Delta_m(r_i)$. The first loss generalizes from $\mathcal{L}_{score}$ to model the outcomes directly
\begin{equation}
    \mathcal{L}_{out} = -\frac{1}{N}\sum_{i,y} 1_i^{(y)}\omega_i^{(y)} \left(m_i\log \mu_i^{(y)} + (1-m_i)\log(1-\mu_i^{(y)})\right),
\end{equation}
where $1_i^{(y)}=I[y_i=y]$ is to mask out unobserved outcomes. We can apply similar methods to the propensity head to compute a propensity loss
\begin{equation}
    \mathcal{L}_{prop} = -\frac{1}{N}\sum_i\left(y_i\log e_i + (1-y_i)\log (1-e_i)\right),
\end{equation}
which is used to calibrate $e_i$ and can be dismissed when we have controlled data distribution with a pre-defined constant $e_i$. A useful regularizer is
\begin{equation}
    \mathcal{L}_{reg} = \frac{1}{N} \sum_{i=1}^N\left(\tau_i-\left(\mu_i^{(1)}-\mu_i^{(0)}\right)\right)^2,
\end{equation}
which encourages the uplift head to agree with the difference of outcome heads, stabilizes the training and improves extrapolation in sparse regions by reducing drifts among heads and sharing statistical strength.

Besides the above three components, we also introduce the DR uplift loss as an auxiliary task to further improve the model performance:
\begin{equation}
    \mathcal{L}_{DR} = \frac{1}{N}\sum_{i=1}^N \omega_i^{DR}\left(\tau_i - \psi_i\right)^2,
\end{equation}
which computes the MSE loss between the uplift and the DR pseudo-outcome
\begin{equation}
    \psi_i = \mu_i^{(1)} - \mu_i^{(0)} + \frac{y_ir_{1,i}}{e_i} - \frac{(1-y_i)r_{0,i}}{1-e_i}
\end{equation}
with a Switch-DR style truncation
\begin{equation}
    \omega_i^{DR} = I\left(\left|\frac{y_ir_{1,i}}{e_i} - \frac{(1-y_i)r_{0,i}}{1-e_i}\right| \leq \lambda_{SW}\right),
\end{equation}
where the residuals are defined as $r_{y,i}=m_i-\mu_i^{(y)}$ and $\lambda_{SW}$ controls the truncation. This helps stabilize the training as the DR signal is unbiased for uplift if either the outcome heads or the propensity is correctly specified (single-robustness on either block). The truncation term improves optimization stability without introducing large bias when outcome heads are reasonable. In practice, we clip the propensity head to avoid divergence of the loss. In this work, we set standard values $\lambda_{SW}=0.01$ and keep it fixed across model variants.

The total loss then becomes
\begin{equation} \label{equ:total_loss}
    \mathcal{L}_{tot} = \lambda_{out} \mathcal{L}_{out} + \lambda_{prop} \mathcal{L}_{prop} + \lambda_{reg} \mathcal{L}_{reg} + \lambda_{DR} \mathcal{L}_{DR},
\end{equation}
where $\lambda$ are loss weights. Unless otherwise noted, we set $\lambda_{out}=\lambda_{prop}=\lambda_{reg}=\lambda_{DR}=1$ in all experiments. We found the outcome loss to dominate naturally, and further tuning of these loss weights did not yield material offline improvements.

An important remark is that the full loss can often be simplified significantly if we only use a designed and balanced data log from randomized traffic. In such a case, we only have real events and the gourd-truth of propensity remains a constant. There’s no statistical reason to estimate the propensity head, and we can assign the known probability in the loss computation and skip $\mathcal{L}_{prop}$. Nevertheless, this head can still be retained for future flexibility and potential imbalance from system instabilities and failures.

Last but not least, we also expand the multi-tasking causal learning schema by introducing multiple $m^{h}$ to further facilitate the learning via shared knowledge between different events and sessions. The above formula can be easily generalized along this dimension, so we'll not discuss the details. One remark is that we do up-weight certain events (on top of the weight discussed above) to ensure model performance on some sparse events and to align better with our business objectives. Thus, the final loss of the model reads
\begin{eqnarray} \nonumber
    \mathcal{L}_{final} &=& \frac{1}{H}\sum_{h=1}^{H}\omega^h \left(\lambda_{out} \mathcal{L}_{out}^{m^{h}}+ \lambda_{reg} \mathcal{L}_{reg}^{m^{h}} + \lambda_{DR} \mathcal{L}_{DR}^{m^{h}})\right) \\
    &+& \lambda_{prop} \mathcal{L}_{prop},
\end{eqnarray}
where we apply constant weights $\omega^h$ for a total of $H$ metrics and $\mathcal{L}_{out}^{m^{h}}$ denotes the outcome loss evaluated on the metric $m^{h}$ only. The propensity loss remains unweighted for different events.

\begin{algorithm}
  \caption{Offline replay evaluation}
  \label{alg:offline_replay}
  \begin{algorithmic}[1]
    \State Input: Offline eval table $T_e$, each record contains fields $p(m\mid y,r)$, $observedMetrics$ and $isTriggered$.
    \State Output: A map from thresholds to goal metrics
    \State
    \State $ \textit{requestCount} \gets 0$, $ \textit{metrics} \gets [\,]$, \textit{results} $\gets$ \{\}
    \State  \textit{thresholds} $\gets$ \textit{np.linspace}(\textit{minThreshold}, \textit{maxThreshold}, \textit{numEvalThresholds})
    \State Sort $T_e$ by $p(m \mid y,r)$
    \State
    \For{each $isTriggered$, $observedMetrics$ in $T_e$}
        \If{not $isTriggered$}
            \State $metrics$ += $observedMetrics$
            \State $requestCount$ += 1 \Comment{For triggering rate eval}
        \EndIf
    \EndFor
   \State \textit{results}[$minThreshold$] = ($requestCount$, $metrics$)
    \State
    \State $cursor \gets 0$
    \For{$\delta_c \in \textit{thresholds}$[1:]}
      \While{$cursor < len(T_e)$ and $\Delta[cursor] <= \delta_c$}
        \State $isTriggered$,  $observedMetrics$ $\gets T_e[cursor]$
        \State $sign \gets int(isTriggered)*2-1$ 
        \State $metrics$ += $sign*observedMetrics$
        \State $requestCount$ += $sign$
        \State $cursor$ += 1
      \EndWhile
      \State \textit{results}[$\delta_c$] = ($requestCount$, $metrics$)
    \EndFor
    \State \Return $results$
  \end{algorithmic}
\end{algorithm}

\subsection{Model architecture}
The model consumes features from the request $r_u$ (e.g., users' demographic information, engagement history and interests), $r_c$ (e.g., time of request, entry point and platform) and $r_q$ (e.g., visual and text embeddings of the query if available). To keep the serving easier and to make the model lightweight since it's on the top of the funnel, we use lots of pretrained visual, text, graph and user sequence embeddings instead of consuming raw features directly \cite{PinsItemSage, PinsLinkSage, PinsPinSage}.

All sparse features are cast into dense embeddings via separate lookup tables and then grouped before concatenating. We use a DCNv2 for feature crossing as well as an MMoE module to accommodate multi-tasking learning.

\subsection{Offline metrics and replay}
To evaluate the model reliably offline, we use both ROC AUC and PR AUC since the events can be sparse. While ROC AUC evaluates the model's capability to rank positives higher than negatives, it's not sensitive to very imbalanced data. PR AUC is more suitable for our task, but we are also interested in predicting the requests without commerce actions accurately. In this sense, we can use a Reverse PR AUC (R-PR AUC), which is computed with flipped ground-truth labels and the same logits, as an additional evaluation metric. We also use Precision@K (P@K) and Recall@K (R@K) for more granular assessment of the model, and similarly, we can define their reverse metrics: Reverse Precision@K (R-P@K) and Reverse Recall@K (R-R@K).

Besides the per-head evaluation metrics, it's also important to determine $\delta_c(r)$ through offline replay. We use the logged randomized holdout data to simulate what happens under a candidate threshold. The logged data contains both treatment arms: requests for which shopping CGs were triggered and requests for which they were not triggered. Given an arbitrary threshold, the policy determines whether each request $r_i$ should trigger shopping CGs. A request is then selected for final metric evaluation only if the policy decision coincides with its logged treatment assignment.

This replay can be viewed as interpolation between the two extreme logged policies in policy space. Therefore, the method does not assume that business metrics are linear or convex functions of the threshold. Its validity instead relies on standard overlap and randomization assumptions.
%: because the holdout assigns $y=0$ and $y=1$ with equal probability, the treated and untreated subsets are comparable in expectation for requests with the same feature distribution.
We also assume no interference between replayed units. In our setting this assumption is structurally mitigated, though not fully eliminated, because the triggering decision is made at the request level before retrieval, ranking and blending, limiting cross-request interactions during the replayed decision itself.

A possible implementation is presented in Algo.~\ref{alg:offline_replay}, where we assume a DP model on a single goal metric $m$. The design has been optimized to have a linear time complexity of $\mathcal{O}(n_e)$, where $n_e$ is the total number of records for evaluation, and can be easily generalized to more complicated policies. The algorithm outputs a set of goal metrics on different thresholds, which is used to determine the actual thresholds for online serving and more details with case studies are presented in Sec.~\ref{sec:experiments}.

\section{Experiments} \label{sec:experiments}
We carry out the experiments of the proposed deep-learning causal retrieval optimization framework in the Pinterest Closeup surface, where related contents are recommended when a user closeups (CUs, i.e., clicks) on a Pin. This surface is critical to guide users through the shopping journey from inspiration to realization with its unique rabbit-hole experience and has many high-quality shopping CGs for driving effective e-commerce distribution, making it ideal to test the proposed framework.

\subsection{Experiment setup}
\subsubsection{Training and evaluation datasets} We use the data discussed in Sec.~\ref{sec:data_logging} for both training (14 days) and evaluation (1 day). While the shopping event is generally sparse, we sample the dataset by certain segments with more e-commerce activities, like country, as well as user and Pin interests (e.g., a fashion Pin is usually considered as more relevant to shopping). This leads to an increase from 1\% to around 10\% of shopping impressions and engagements in the final training data, helping mitigate the data sparsity issue. 

\begin{table}[h]
  \caption{Positive rate of different metrics $m$ in final training data segmented by the treatment.}
  \label{tab:event_stats}
  \begin{tabular}{ccccc}
    \toprule
    Metrics $m$ & Total & $y=0$ & $y=1$ & Lift  \\
    \hline
    Success Curation & 34.09\% & 34.52\% & 33.64\% & -2.62\% \\
    Mid Funnel & 34.77\% & 35.2\% & 34.32\% & -2.56\% \\
    Lower Funnel & 2.69\% & 2.55\% & 2.84\% & +10.21\% \\
    Shopping Top Funnel & 18.81\% & 18.36\% & 19.27\% & +4.72\% \\
    Shopping Mid Funnel & 13.35\% & 13.28\% & 13.48\% & +1.12\% \\
    Shopping Lower Funnel & 1.38\% & 1.14\% & 1.61\% & +29.19\% \\
    Repins & 34.42\% & 34.87\% & 33.97\% & -2.65\% \\
    Long-clicks & 1.59\% & 1.52\% & 1.65\% & +7.88\% \\
    Shopping repins & 1.86\% & 1.5\% & 2.22\% & +32.43\% \\
    Shopping long-clicks & 0.71\% & 0.59\% & 0.83\% & +28.92\% \\
    \bottomrule
  \end{tabular}
\end{table}

\subsubsection{Goal metrics and coverage} In Tab.~\ref{tab:event_stats}, we summarize the coverage of goal metrics $m^{(h)}$ of interests and their lift from the ablated group ($y=0$). They can be divided into two types, and the first one is the common engagements. While there are tons of different types of engagements on the Pinterest platform, we focus on repin (RP, when a user saves a Pin to a board) and long-click (LC, when a user clicks on the link of the Pins and spends more than 35 seconds) and their segmented version on shopping events.

From the results, we can see that the shopping CGs contribute significantly to both shopping RPs and LCs with a lift of around 30\% while it reduces the overall RPs by -2\%. One reason is that when users LC, they will directed to an external website and they might not come back, leading to fewer engagements on the platform. Another possible interpretation is that overloading shopping contents uniformly to all users leads to a less satisfactory experience.

The other type of metrics of particular interest is session-related, and they are segmented by shopping (e.g., shopping lower funnel or SLF) and overall (e.g., mid funnel) sessions similarly. Those session metrics are typically associated with certain types of events, for example, the top funnel is usually related to impressions, the mid funnel depends on RPs and other social events, and finally, the lower funnel heavily counts LCs. The shopping CGs can lift SLF sessions proportionally to shopping LCs, and similarly, degrade the overall mid-funnel session. However, we do notice that the lift of shopping mid funnel (SMF) is much less, indicating the shopping CGs are mostly driving users RP more in a single session. In other words, most of the mid funnel can already be satisfied by other non-shopping CGs.

\subsubsection{Model and training hyperparameter}
For model training, we select all ten events in Tab.~\ref{tab:event_stats}, along with a negative event and a refined version of the SLF session. The default head weight is set to $\omega^h=1$, and we up-weight it for sparse events like LCs and lower-funnel sessions with $\omega^h=10$. This event-level upweighting reflects business prioritization and label sparsity rather than delicate optimization (usually consistent with downstream models). We randomly select 5M records from the training datasets to compute the feature statistics, which are later used to initialize the embedding lookup tables and to normalize the input features. 

For later convenience, we define a baseline model trained using $\mathcal{L}_{score}$. This model has a 3-layer DCNv2 module and an MMoE module with 12 experts, and each contains a [256, 64, 32] hidden layer. As a result, the baseline model has around 170M total trainable parameters. For models trained with the DR learning framework, we fix the weights of the loss functions to be 1 in Equ.~\ref{equ:total_loss}, and fine-tuning doesn't lead to better performance since the outcome loss is often dominant under the controlled holdout design.

The model is set up as an instance of PyTorch DistributedDataParallel class, trained on an AWS p4d.24xlarge instance with 8 NVIDIA A-100 GPUs and optimized with an FuseAdam optimizer with 0.001 learning rate and 0.9/0.999 betas. We choose a batch size of 8192 to ensure a more stable and uniform in-batch distribution of the sparse events. Our offline results indicate that the model doesn't benefit significantly from a larger batch size or fine-tuning of the optimizer, given its lightweight nature.

\subsubsection{Online serving} To prepare for online serving, the model is converted to TrochScript and then deployed to our in-house inference service Scorpion as depicted in Fig.~\ref{fig:sys_overview}. This model is called on-the-fly with more than 100k peak QPS to determine if the current request should trigger the shopping CGs by checking the given policy. This model is also called in parallel with other remote procedure calls, like the inference of user tower in the two-tower based CGs, to save time, and thus, it doesn't lead to any materialized increases of the end-to-end serving latency, eliminating potential noises to the online experiments.

\begin{table}[h]
  \caption{Offline evaluation of AUC of key session metrics in the evaluation dataset.}
  \label{tab:offline_eval}
  \begin{tabular}{ccccc}
    \toprule
     & 0.75 epoch & 1.25 epoch & Larger & DR + Larger  \\
    \hline \hline
    ROC AUC &  &  &  &  \\
    \hline
    MF & +0.02\% & -0.30\% & +0.00\% & -0.52\% \\
    LF & -0.01\% & -1.32\% & -0.02\% & -0.21\% \\
    SMF & +0.03\% & -4.90\% & -0.02\% & -0.25\% \\
    SLF & -0.03\% & -2.20\% & -0.04\% & -0.14\% \\
    \hline \hline
    PR AUC &  &  &  &  \\
    \hline
    MF & -0.12\% & +3.35\% & +0.05\% & -1.16\% \\
    LF & -0.35\% & -0.17\% & -0.64\% & -4.19\% \\
    SMF & -0.10\% & +3.24\% & +0.27\% & -1.98\% \\
    SLF & -1.86\% & +0.45\% & -2.89\% & -4.29\% \\
    \hline \hline
    R-PR AUC &  &  &  &  \\
    \hline
    MF & -0.00\% & -7.90\% & -0.01\% & +0.33\% \\
    LF & +0.01\% & -3.24\% & +0.07\% & +0.07\% \\
    SMF & -0.01\% & -10.80\% & -0.00\% & +0.14\% \\
    SLF & +0.01\% & -4.02\% & +0.02\% & +0.04\% \\
    \bottomrule
  \end{tabular}
\end{table}

\subsection{Offline evaluation and optimization}
\subsubsection{Per-head prediction performance} We start by looking at the prediction performance of each head using the baseline model described above and some fine-tuned parameters as discussed in previous sections. Some offline evaluation results are summarized in Tab.~\ref{tab:offline_eval}, where we compare the baseline models with models trained on different epochs, a larger model with more trainable parameters (Larger group, around 5x of the baseline) and one larger model that's trained with the DR loss $\mathcal{L}_{final}$ (DR + Larger group).

We first notice that the common over-fitting after 1 epoch in the recommendation system is also present in this model. This can be ascribed to some shared traits with other recommendation models (e.g., the ranking model), like data sparsity, data shifting pattern and large embedding tables. In light of this, we used an L2 regularization during the training, but it doesn't bring significant improvements to the model's generalization capability.

Another observation is that simple scaling up of the model size doesn't naturally lead to AUC gains. This can be attributed to multiple factors, including model architecture and training data (e.g., a more scalable architecture like transformers with raw user sequence features might help). But most importantly, the AUC metrics already saturate since the base model actually has a 0.97 ROC AUC, a 0.35 PR AUC and a 0.95 R-PR AUC for MF. As a result, AUC is not sensitive to the local changes that matter most for the triggering decision.

This distinction is important because our production policy does not use the full ranking distribution uniformly. Instead, it operates in a high-precision threshold regime, where the model must identify requests that should be confidently triggered or confidently suppressed. Therefore, standard AUC can understate improvements that are concentrated near the operating threshold. We consequently use precision P@K, recall R@K and their reverse metrics for a more decision-aligned evaluation, as summarized in Tab.~\ref{tab:offline_eval_TPLF}. These metrics also help distinguish a truly better triggering policy from a larger model that simply over-triggers shopping CGs.

\begin{table}[h]
  \caption{Offline evaluation of (R-)P@K and (R-)R@K of SLF in the evaluation dataset.}
  \label{tab:offline_eval_TPLF}
  \begin{tabular}{ccccc}
    \toprule
    & P@10 & P@20 & P@50 & P@100  \\
    \hline
    % 0.75 epoch & +12.50\% & +6.25\% & +11.11\% & -1.33\% \\
    % 1.25 epoch & -12.50\% & -31.25\% & -16.67\% & -9.33\% \\
    Larger      & -12.50\% & -12.50\% & +2.78\%  & -8.00\% \\
    DR loss   & -12.50\%   & -6.25\%    & +5.56\%    & -5.33\%     \\
    DR + Larger      & +12.50\% & +18.75\% & +22.22\% & +1.33\% \\
    \hline \hline
    & R-P@10 & R-P@20 & R-P@50 & R-P@100  \\
    \hline
    % 0.75 epoch & -50.00\% & -25.00\% & -28.57\% & +4.00\% \\
    % 1.25 epoch & +50.00\% & +125.00\% & +42.86\% & +28.00\% \\
    Larger      & +50.00\% & +50.00\%  & -7.14\%  & +24.00\% \\
    DR loss   & +50.00\%   & +25.00\%   & -14.29\%   & +16.00\%    \\
    DR + Larger      & -50.00\% & -75.00\%  & -57.14\% & -4.00\% \\
    \hline \hline
    & R@10 & R@20 & R@50 & R@100  \\
    \hline
    % 0.75 epoch & +12.41\%  & +0.00\%   & +9.80\%   & +9.48\% \\
    % 1.25 epoch & +162.53\% & +113.95\% & +216.18\% & +245.47\% \\
    Larger   & -12.56\%    & -12.50\%   & +2.79\%     & -8.01\%     \\
    DR loss   & -12.56\%    & -6.23\%    & +5.56\%     & -5.34\%     \\
    DR + Large& +12.44\%    & +18.81\%   & +22.20\%    & +1.33\%     \\
    & R-R@10 & R-R@20 & R-R@50 & R-R@100  \\
    \hline
    % 1x25\_epoch & +733.33\% & +1133.33\% & +650.00\% & +576.92\% \\
    % 0x75\_epoch & -33.33\% & -16.67\% & -31.82\% & +2.56\% \\
    Larger      & +66.67\% & +50.00\% & -9.09\% & +23.08\% \\
    DR loss   & +66.67\%    & +33.33\%     & -13.64\%     & +15.38\%    \\
    DR + Larger      & -33.33\% & -66.67\% & -59.09\% & -5.13\% \\
    \bottomrule
  \end{tabular}
\end{table}

Because SLF is more relevant to the following discussions, we only present the results of this head in Tab.~\ref{tab:offline_eval_TPLF}. For the standard metrics, the DR + Larger group significantly outperforms both the baseline and the Larger group across almost all K values. For the reverse metrics, the DR + Larger group also shows sizable reductions, meaning it is less likely to push shopping content to users who do not like it.

These results indicate that model capacity alone is insufficient and can even amplify mis-targeted triggering. The DR objective appears to channel the additional capacity toward better separation of requests with and without incremental shopping value. Overall, the combination of DR loss and larger capacity is what yields both higher recall of valuable shopping opportunities and lower exposure to users unlikely to benefit from shopping content.

\begin{figure}[h]
  \centering
  \includegraphics[width=\linewidth]{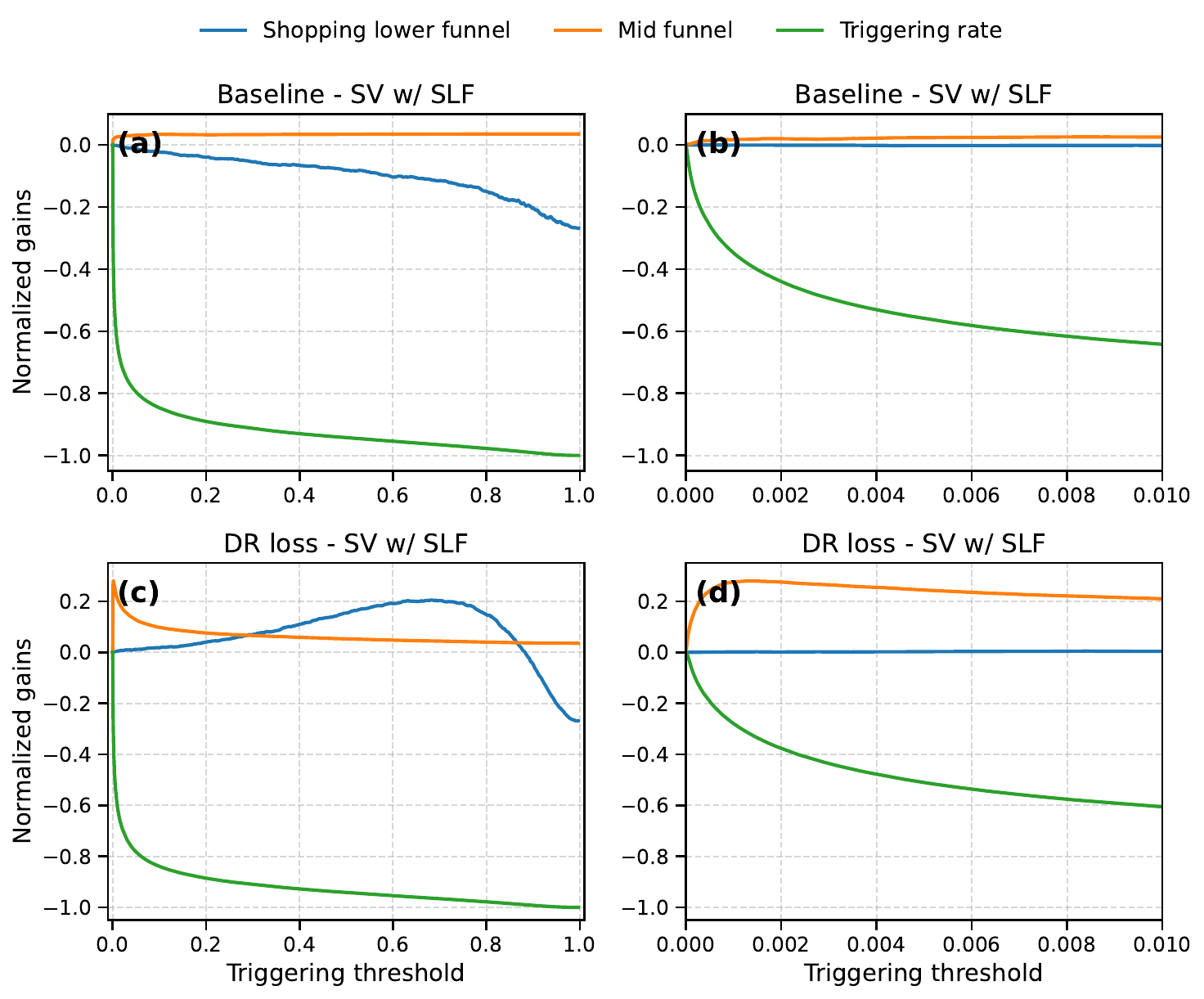}
  \caption{Offline replay of SV with the SLF head.}
  \label{fig:sv_slf}
\end{figure}

\subsubsection{Offline replay} We further run the offline replay proposed in Algo.~\ref{alg:offline_replay} on the DR loss group and compare it with the baseline model in Fig.~\ref{fig:sv_slf} using the SV policy, where only the $y=1$ prediction is used to control the triggering. In the plot, all the metrics are normalized as the relative gains to $\delta_c=0$, so they must be zero when $\delta_c=0$ and reflect the lift in Tab.~\ref{tab:event_stats} when $\delta_c=1$.

First and foremost, an exponential drop in triggering rate and a linear decay of SLF are observed in Fig.~\ref{fig:sv_slf}(a), which suggests that the model is able to select the right requests for shopping content distribution. Meanwhile, with fewer requests triggering shopping CGs, the mid funnel slowly recovers, indicating that the model could help improve the mid funnel with exponentially fewer triggering of shopping CGs but a relatively small tradeoff of SLF. Such a critical exponential versus linear behavior is clearer when we zoom in on the small $\delta_c$ region in Fig.~\ref{fig:sv_slf}(b).

We now move to examine the DR loss group, which exhibits quite different behaviors as shown in Figs.~\ref{fig:sv_slf}(c) and (d). With a small $\delta_c$, the triggering rate drops quickly, while we observe a huge improvement in the mid funnel as well. In the meantime, the SLF remains constant, meaning that the model can achieve great gains in driving the mid-funnel events without any tradeoffs in the SLF. The mid funnel soon reaches a peak before it starts to drop slowly, while the SLF initially increases - likely a result of more efficient targeting at users who are interested in commerce content. But eventually, it hits a critical point where we have to take tradeoffs in SLF if we want to trigger even less. From the results, it's clear that the DR loss group show much better performance as it can identify users both with and without shopping intents.

\begin{figure}[h]
  \centering
  \includegraphics[width=\linewidth]{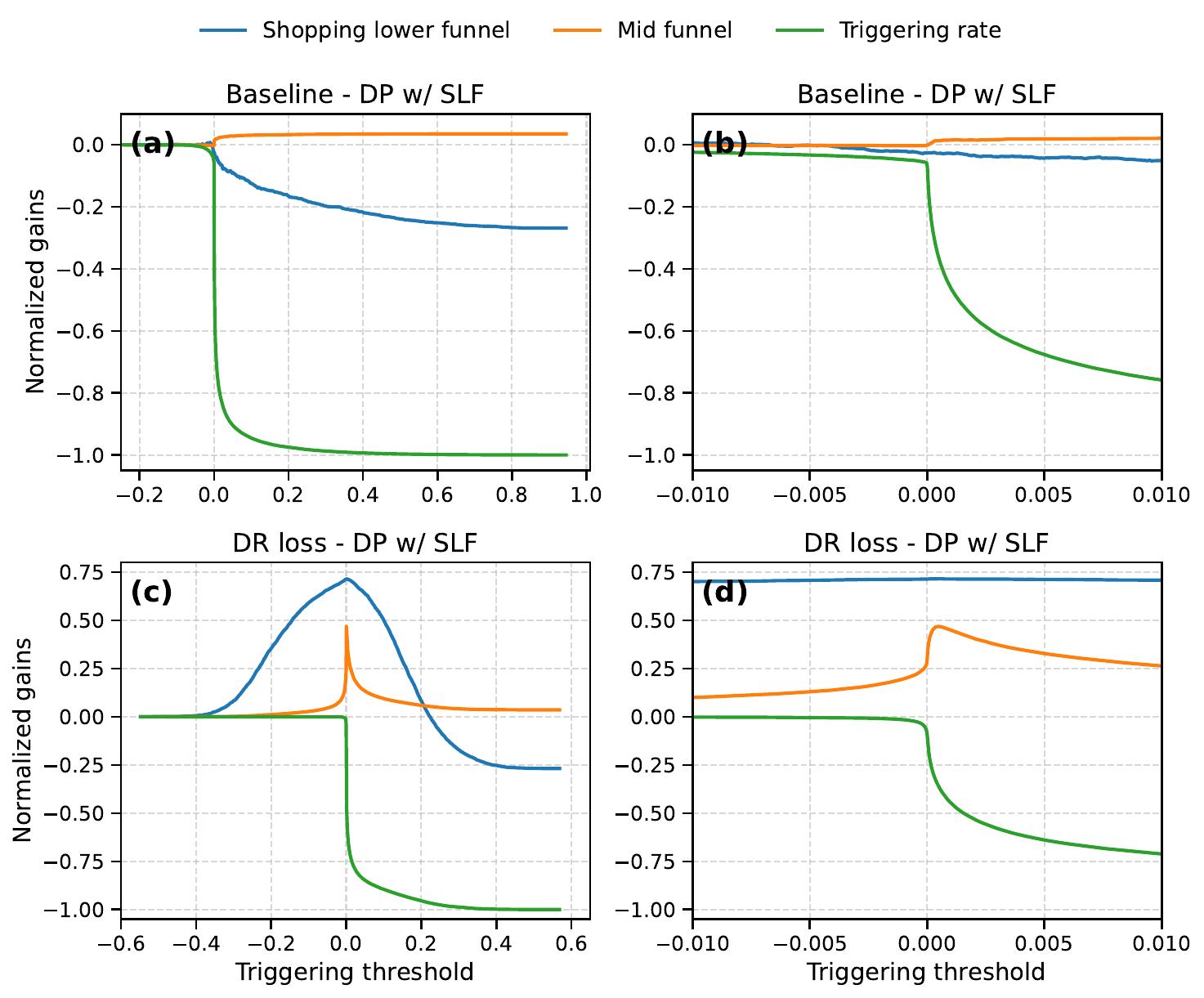}
  \caption{Offline replay of DP with the SLF head.}
  \label{fig:dp_slf}
\end{figure}

Similar results using DP are presented in Fig.~\ref{fig:dp_slf}. We start with the baseline model in panels (a) and (b), and now the critical behaviors happen around $\delta_c=0$ as expected. Once crossing the point, the triggering rate drops exponentially, and the mid funnel shows a small jump while SLF remains almost linear with a small negative slope. The DR loss group again exhibits quite different patterns as shown in Figs.~\ref{fig:dp_slf}(c) and (d). With increasing threshold from left to right, the SLF starts to increase first and then drops. At the critical point $\delta_c=0$, mid funnel shows a sharp and asymmetric peak while the SLF hits a narrow plateau (see Fig.~\ref{fig:dp_slf}(d)). The above results provide strong support that the multi-tasking DR loss improves targeting accuracy and reduces mis-targeted exposures, which combine to maximize the effect of shopping triggering.

% \begin{figure}[h]
%   \centering
%   \includegraphics[width=\linewidth]{model_comparsion.pdf}
%   \caption{Offline replay comparing different groups in Tab.~\ref{tab:offline_eval_TPLF}.}
%   \label{fig:dp_slf}
% \end{figure}

\subsubsection{Feature importance analyses} We use permutation-based approaches to evaluate the feature importance to gain more interpretability of the model's triggering decisions. Below, we discuss some of the top features across all the models.

For user-side features, the most important ones are user representations learned from next-action loss on the user sequence. Besides the user's on-site shopping activities, their off-site engagements are often ranked as top 3 since they cater to the user's interests in shopping outside our platform. Some other very important user features include country and language, since our inventory may have higher-quality shopping content from certain countries. Simple counting of user engagements like CUs is also ranked in the top 10.

Besides user features, we find that two context features are often ranked in the top 5, including the entry point of the Closeup surface and the platform (or app type). The former is related to the observation that users coming from the Search Feed have stronger intent than those from the Home Feed. In terms of the platform, a hypothesis is that users are more comfortable with commerce activities on certain devices.

For the query Pins, their embeddings jointly learned from the user representations discussed above are usually ranked higher. Other top-ranked ones are multi-modal embeddings like ItemSage \cite{PinsItemSage} and sparse features like style vectors that could help the model infer shoppable contents in the Pin. Counting features like numbers of Imps, CUs and RPs/LCs are also relatively important since they reflect the overall quality and popularity of the Pin.

Such a result suggests that the model can combine the user's shopping intent with high-quality shoppable content of the Pin under the right context to predict both the outcomes and uplift. This in turn validates the effectiveness of the proposed learning objective, model architecture and training methods.

\begin{table}[h]
  \caption{Offline replay versus online for SV policy using the baseline model.}
  \label{tab:single_value_offline_online}
  \begin{tabular}{ccccc}
    \toprule
    $\delta_c$ & \multicolumn{2}{c}{Triggering rate} & \multicolumn{2}{c}{TP Imp} \\
    & Online & Offline & Online & Offline \\
    \hline
    0.0383 & -77.4\% & -78.5\% & -19.3\% & -18.1\%  \\
    0.0179 & -70.4\% & -71.3\% & -14.3\% & -13.6\% \\
    0.0102 & -64.4\% & -64.6\% & -11.3\% & -10.1\% \\
    0.0055 & -57.0\% & -57.1\% & -8.0\% & -7.6\% \\
    0.0023 & -46.1\% & -46.0\% & -5.0\% & -4.9\% \\
    0.0012 & -37.0\% & -37.5\% & -2.3\% & -3.3\% \\
    \bottomrule
  \end{tabular}
\end{table}

\subsection{Online experiments}
We brought the promising groups from offline assessment to online experiments. First of all, we validate the offline replay results by comparing the actual triggering rate, TP Imp and TPLC with the prediction obtained in previous sections. The summary in Tab.~\ref{tab:single_value_offline_online} indicates a surprising consistency between online observation and offline evaluation, which is usually hard to achieve in the early-retrieval stage due to downstream and systematic effects. The fidelity of replay is also metric-dependent. For policy-controlled upstream metrics such as triggering rate and impressions, replay is nearly exact because these quantities are directly determined by the triggering decision. For mid-funnel and engagement-level metrics such as clicks, replay remains directionally reliable but accumulates additional variance from retrieval, ranking and blending. For downstream session metrics, replay should be interpreted primarily as a trend estimator rather than an exact point predictor. This level of fidelity is sufficient for threshold selection before launch, while final validation still relies on online experiments. The consistency is also robust since we observe it with other models like the DR loss group.

However, for DP, the metric consistency is usually low, possibly due to the high variance of the uplift as discussed in Sec.~\ref{sec:problem_def}. We do note that variance-reduction techniques for DP, such as ensembles, uncertainty-aware filtering near the boundary, or more targeted exploration in sparse regions, are promising future directions.

\begin{table}[h]
  \caption{Online metric wins.}
  \label{tab:online_metrics}
  \begin{tabular}{ccc}
    \toprule
    Goal Metric & Volume & Propensity \\
    \hline
    Success Curation (site-wide) &  +0.26\% & nss \\
    Mid funnel (site-wide) & +0.46\% & +0.21\% \\
    Shopping lower funnel (site-wide) & nss & nss \\
    Session > 5 min (site-wide) & +0.46\% & nss \\
    Repins (Closeup surface) & +1.10\% & +0.43\% \\
    \bottomrule
  \end{tabular}
  
  \vspace{0.5ex}
  {\footnotesize\textit{Note.} nss stands for not statistically significant.\par}
\end{table}

The key business metrics from the best-performing online group are summarized in Tab.~\ref{tab:online_metrics}. As expected from the offline replay, we see recovery of both overall and mid-funnel sessions with neutral SLF. In this specific group, the online triggering of shopping CGs is reduced by around 40\%, which leads to an extra saving of nearly half a million after subtracting the serving cost. In one of the most aggressive groups, we reduced the triggering rate by up to 85\% and observed neutral SLF sitewide.

\section{Discussion}
A deliberate production choice in this work is to decouple model learning from final policy selection. The model learns outcome and uplift signals, while offline replay and threshold selection convert these signals into a deployable policy. Although this means the final policy is not learned fully end-to-end, the decoupling provides important production advantages by making the tradeoff curve auditable, allowing product and infrastructure stakeholders to inspect the effect of each threshold, and enabling threshold updates without retraining the model when business priorities shift. In early-retrieval systems, this explicit control is often preferable to a fully automated objective whose behavior may be harder to negotiate or debug across teams.

The proposed framework can be generalized beyond binary triggering $y$ and binary metric $m$ by introducing them in the continuous space to accommodate broader use cases and more granular control. A common approach to extend the binary triggering is to bucketize the fetch count into zero, small, medium and large fetch counts.  Some important metrics can be modeled beyond the direct business metrics discussed so far, e.g., infra/cost budgets, enriching the whole framework for a holistic early-retrieval optimization.

While in this work, we focus on discussing the SLF head for triggering, as it's the single most relevant metric to our business, this framework can be extended to triggering via multiple $m^i$. Two straightforward ways are using logic operation of single-head policy like $\lor_i \Delta_{m^i}>\delta_{c,i}$ and using a utility function as typical in post-ranking or blending stage $\left(\sum_i \omega_im^i\right)>\delta_c$, where $\omega_i$ are tunable weights. Both offline reply and online serving can be extended accordingly, providing flexibility to handle more complicated business requirements.

This design can also support multiple CGs, given diversity at the retrieval stage. However, when combining with policies beyond binary, this brings fundamental challenges with an exponentially growing exploration space $n_p^{N_r}$, where $n_p$ is the number of policies for a single CG and $N_r$ is the total number of CGs. Several approaches can be designed to trim the exploration space, including removing invalid policies and introducing stratified sampling (target exploration) to focus on policies with desired business impacts while allowing some levels of exploration of more general policies. Besides, reinforcement learning (RL) approaches can often bring additional value in those cases, and we'd like to point out some similarities between our proposal and value-based RL, like deep Q-network. Specifically, our framework uses the uplift $\tau$ as the reward function (for DP methods), the whole user request $r$ as the state and $y$ as the action. 

For smaller platforms or surfaces with less randomized traffic, the same framework may require additional data-efficiency measures. Possible mitigation includes using propensity-score matching on historical logs to construct quasi-counterfactual pairs, running short targeted A/B tests to collect high-value exploration data, pretraining outcome heads on larger observational datasets before DR fine-tuning, or replacing rare downstream labels with denser proxy metrics during early iterations. These variants trade off statistical purity, engineering cost and launch risk, but they preserve the main principle of learning trigger policies from incremental rather than purely correlational signals.

\begin{table}[h]
  \caption{Triggering rate versus threshold for SV policy trained on different dates and loss.}
  \label{tab:model_stability}
  \begin{tabular}{cccc}
    \toprule
    $\delta_c$ & Baseline model & Baseline model & DR loss model \\
     & Feb 2025 & Oct 2025 & Oct 2025 \\
    \hline
    0.050 & -78.51\% & -79.59\% & -80.92\%  \\
    0.020 & -71.30\% & -71.48\% & -72.66\% \\
    0.010 & -64.67\% & -64.23\% & -65.22\% \\
    0.005 & -57.11\% & -55.98\% & -56.79\%  \\
    0.002 & -46.01\% & -44.09\% & -44.64\% \\
    0.001 & -37.50\% & -35.00\% & -35.31\% \\
    \bottomrule
  \end{tabular}
\end{table}

To stably serve this model in production and keep regular model refreshing and iteration, the model must be robust to the evolving ecosystem, serving failures and user-behavior drift. In light of this, we summarize in Tab.~\ref{tab:model_stability} how the triggering rate varies with respect to $\delta_c$ across models trained on different dates and with different losses. Quite surprisingly, although those models can lead to different engagement metrics, the triggering-rate curves are highly consistent despite changes in the retrieval layer and downstream ranking/blending systems.

A reasonable hypothesis is that the model captures intrinsic shopping intent, which is less affected by short-term recommendation-system changes and evolves more slowly over time. To validate this, we further performed an ablation of feature groups from the request $(r_u, r_c, r_q)$. When either context $r_c$ or query $r_q$ features are removed, we see small and comparable deviations. However, the gap becomes large if we ablate the user feature group $r_u$. This indicates strong personalization in the model predictions and suggests that user-side intent signals are the main driver of policy stability.

We nevertheless emphasize that the causal replay relies on assumptions that may degrade under larger distribution shifts. In particular, the method assumes overlap from the randomized holdout, no severe interference between replayed units, and sufficient similarity between holdout traffic and future production traffic. The request-level triggering design mitigates some interference because the decision is made before retrieval, ranking and blending, but it does not formally eliminate all downstream interactions. In practice, we address behavior drift through periodic model refreshes, continuous monitoring of triggering-rate curves, and covariate-drift checks between live traffic and holdout data.

Another important topic in production is the potential feedback loop. Besides long-term feedback loops, which make it longer for the system to converge after launch, there are also more real-time feedback loops on user experience. A possible scenario is when the model relies heavily on near-term shopping engagements to make the triggering decision. While the exact characterization of the feedback loop, e.g., using Markovian process modeling, is beyond the scope of this work, it is important for future analyses. During our experiments, we observed that the actual online triggering rate drifts down slowly in month-over-month comparisons, yet this did not lead to significant shifts in engagement metrics and no detrimental effects were observed.

Finally, robustness to noisy and incomplete data is a central production requirement. The proposed system includes several safeguards. First, the propensity head is retained even though the main holdout has a known constant propensity, making the framework more flexible if future logging deviates from the controlled 50/50 design. Second, multi-task learning transfers statistical strength from dense engagement heads to sparse lower-funnel heads, while event-level upweighting prevents rare but business-critical events from being ignored. Third, Switch-DR truncation removes records with extreme inverse-propensity corrections from the uplift loss, preventing unstable gradients. Fourth, reverse metrics such as R-PR AUC and R-P@K explicitly monitor over-triggering and mis-targeted exposure. While we do not provide a full quantitative uncertainty estimator in this work, adding ensemble-based confidence intervals or conformal filtering around the decision boundary is an important future direction, especially for more aggressive DP policies.

\section{Conclusion}
In conclusion, we have proposed a deep-learning-based causal inference framework for CG triggering to optimize the e-commerce distribution in Pinterest. We developed both a sophisticated model and a reliable offline replay to guide model iteration. This leads to neutral shopping sessions and huge improvements in overall engagement with a significant reduction of infrastructure cost. This system has been deployed to the production system to serve billions of requests daily at Pinterest.

\begin{acks}
J. Hou would like to thank K. Wahba, T. Wang, V. Arun for setting up and analyzing the shopping ablation holdout as well as providing the feature importance tool; M. David for insights of DR learning; J. Hsieh and R. Han for important product inputs; N. Tran and Q. Luo for serving supports; W. Li, B. Deng, J. Li, A. Malik, Y. Chen, J. Qu, D. He and W.-T. Lin for helpful discussions across data collection, reinforcement learning, modeling and other approaches.
\end{acks}

\bibliographystyle{ACM-Reference-Format}
\balance
\bibliography{cg_triggering}

% \appendix




\end{document}